\documentclass[aps,prd,superscriptaddress,showpacs,preprint]{revtex4}
\usepackage{graphicx, bm}
\usepackage[usenames]{color}

\begin{document}
%\tightenlines
\draft
\title{Neutral Higgs Boson Pair-Production and Trilinear Self-Couplings in the MSSM
       at ILC and CLIC Energies}

\author{ A. Guti\'errez-Rodr\'{\i}guez}
\affiliation{\small Facultad de F\'{\i}sica, Universidad Aut\'onoma de Zacatecas\\
         Apartado Postal C-580, 98060 Zacatecas, M\'exico.\\}

\author{M. A. Hern\'andez-Ru\'{\i}z}
\affiliation{\small Instituto de F\'{\i}sica, Universidad
        Aut\'onoma de San Luis Potos\'{\i}\\
        78000 San Luis Potos\'{\i}, SLP, M\'exico.\\}

\author{ O. A. Sampayo}
\affiliation{\small Departamento de F\'{\i}sica, Universidad Nacional del Mar del Plata\\
              Funes 3350, (7600) Mar del Plata, Argentina.}

\date{\today}
%\maketitle

\begin{abstract}
% insert abstract here

We study pair-production as well as the triple self-couplings of
the neutral Higgs bosons of the Minimal Supersymmetric Standard
Model (MSSM) at the Future International Linear $e^{+}e^{-}$
Collider (ILC) and Compact Linear Collider (CLIC). The analysis is
based on the reactions $e^{+}e^{-}\rightarrow b \bar b h_ih_i, t
\bar t h_ih_i$ with $h_i=h, H, A$. We evaluate the total
cross-section for both $b\bar bh_ih_i$, $t\bar th_ih_i$ and
calculate the total number of events considering the complete set
of Feynman diagrams at tree-level. We vary the triple couplings
$\kappa\lambda_{hhh}$, $\kappa\lambda_{Hhh}$,
$\kappa\lambda_{hAA}$, $\kappa\lambda_{HAA}$,
$\kappa\lambda_{hHH}$ and $\kappa\lambda_{HHH}$ within the range
$\kappa=-1$ and +2. The numerical computation is done for the
energies expected at the ILC with a center-of-mass energy 500,
1000, 1600 $GeV$ and a luminosity 1000 $fb^{-1}$. The channels
$e^{+}e^{-}\rightarrow b \bar b h_ih_i$ and $e^{+}e^{-}\rightarrow
t \bar t h_ih_i$ are also discussed to a center-of-mass energy of
3 $TeV$ and luminosities of 1000 $fb^{-1}$ and 5000 $fb^{-1}$.
\end{abstract}

\pacs{ 13.85.Lg, 12.60.Jv
\vspace*{2cm}\noindent  E-mail: $^{1}$alexgu@planck.reduaz.mx, }

\vspace{5mm}

\maketitle

\section{Introduction}

The search for Higgs bosons is one of the principal objectives of
present and future high-energy colliders, such as the
International Linear Collider (ILC) \cite{TAbe,Aguilar,Koh,ILC},
Compact Linear Collider (CLIC) \cite{Accomando} and Large Hadron
Collider (LHC) \cite{Spira,Djouadi1}. It has been demonstrated in
Ref. \cite{Weiglein} that physics at the LHC and $e^{+}e^{-}$ ILC
will be complementary to each other in many respects. In many
cases, the ILC can significantly improve the LHC measurements.

The Higgs boson \cite{Higgs,Higgs1,Higgs2,Englert,Guralnik} plays
an important role in the Standard Model (SM)
\cite{Weinberg,Salam,Glashow} because it is responsible for
generating the masses of all the elementary particles (leptons,
quarks, and gauge bosons). However, the Higgs-boson sector is the
least tested in the SM, in particular the Higgs boson
self-interaction. If Higgs bosons are responsible for breaking the
symmetry from $SU(2)_L\times U(1)_Y$ to $U(1)_{EM}$, it is natural
to expect that other Higgs bosons are also involved in breaking
other symmetries. One of the more attractive extensions of the SM
is Supersymmetry (SUSY) \cite{Nilles,Haber,Barbieri}, mainly
because of its capacity to solve the naturalness and hierarchy
problems while maintaining the Higgs bosons elementary.

The theoretical framework of this paper is the Minimal
Supersymmetric extension of the Standard Model (MSSM), which
doubles the spectrum of particles of the SM, and the new free
parameters obey simple relations. The scalar sector of the MSSM
\cite{Gunion,Gunion1} requires two Higgs doublets, thus the
remaining scalar spectrum contains the following physical states:
two CP-even Higgs scalares $(h^0, H^0)$ with $M_h \leq M_H$, one
CP-odd Higgs scalar $(A^0)$ and a charged Higgs pair $(H^\pm)$.
The Higgs sector is specified at tree-level by fixing two
parameters which can be chosen as the mass of the pseudoscalar
$M_A$ and the ratio of vacuum expectation values of the two
doublets $\tan\beta=v_2/v_1$; then, the masses $M_h$, $M_H$ and
$M_{H^\pm}$ and the mixing angle of the neutral Higgs sector
$\alpha$ can be fixed. However, since radiative corrections
produce substantial effects on the predictions of the model
\cite{Li,Gunion2,Gunion3,Gunion4,Okada,Haber1,Ellis,Dress}, it is
necessary to also specify the squark masses, which are assumed to
be degenerated.

In particular, all the triple self-couplings of the physical Higgs
particles can be predicted (at the tree level) in terms of $M_A$
and $\tan\beta$. Once a light Higgs boson is discovered, the
measurement of these triple couplings can be used to reconstruct
the Higgs potential of the MSSM.

The triple Higgs self-couplings can be measured directly in
pair-production of Higgs particles at hadron and high energy
$e^{+}e^{-}$ colliders. In proton collisions at the LHC
\cite{Djouadi1}, Higgs pairs can be produced through double
Higgs-strahlung off $W$ and $Z$ bosons, $WW$ and $ZZ$ fusion, and
gluon-gluon fusion. The triple Higgs boson couplings are involved
in a large number of processes at $e^{+}e^{-}$ collider
\cite{Djouadi2,Djouadi3,Osland0,Boudjema,Djouadi4,Ferrera,Osland,Osland2,Osland1,Abdesslam,Lafaye,Miller,Vernon}:

\begin{eqnarray}
\mbox{double Higgs-strahlung}&:& e^+e^- \to ZH_iH_j \hspace{3mm} \mbox{and} \hspace{3mm} ZAA \hspace{3mm} [H_{i,j}=h, H],  \nonumber \\
\mbox{triple Higgs production}&:& e^+e^- \to AH_iH_j \hspace{3mm}
\mbox{and} \hspace{3mm} AAA, \\
\mbox{$WW$ fusion}&:& e^+e^-\to \bar\nu_e \nu_e H_iH_j
\hspace{3mm} \mbox{and} \hspace{3mm} \bar\nu_e \nu_e AA,\nonumber
\end{eqnarray}

\noindent these three-body processes have been evaluated
extensively. However, the inclusion of four-body processes with
heavy fermions $b$ and $t$ ($e^{+}e^{-}\rightarrow b\bar b h_ih_i,
t\bar t h_ih_i, h_i=h, H, A $) is important in order to know its
impact on three-body channels and also to search for new relations
that may have a clear signature of the Higgs boson production. In
the studied processes, the MSSM Higgs bosons are radiated by
$b(\bar b)$ and $t(\bar t)$ quarks at future $e^{+}e^{-}$
colliders
\cite{A.Gutierrez,A.Gutierrez1,A.Gutierrez2,A.Gutierrez3,A.Gutierrez4,A.Gutierrez5}
with a c.m. energy in the range of 500 to 1600 $GeV$, as in the
case of the ILC \cite{TAbe} and of 3 $TeV$ to the CLIC machine
\cite{Accomando}.

In this paper, we study the pair-production as well as the triple
self-couplings of the Higgs bosons of the Minimal Supersymmetric
Standard Model (MSSM) at the Future International Linear
$e^{+}e^{-}$ Collider (ILC) \cite{TAbe} and Compact Linear
Collider (CLIC) \cite{Accomando}. The analysis is based on the
reactions $e^{+}e^{-}\rightarrow b \bar b h_ih_i, t \bar t h_ih_i$
with $h_i=h, H, A$. We evaluate the total cross-section for both
$b\bar bh_ih_i$, $t\bar th_ih_i$ and calculate the total number of
events considering the complete set of Feynman diagrams at
tree-level. We vary the triple couplings $\kappa\lambda_{hhh}$,
$\kappa\lambda_{Hhh}$, $\kappa\lambda_{hAA}$,
$\kappa\lambda_{HAA}$, $\kappa\lambda_{hHH}$ and
$\kappa\lambda_{HHH}$ within the range $\kappa=-1$ and +2. The
numerical computation is done for the energies expected to be
available at the ILC with a center-of-mass energy 500, 1000, 1600
$GeV$ and a luminosity 1000 $fb^{-1}$. Our analysis is also
extended to a center-of-mass energy 3 $TeV$ and luminosities of
1000 $fb^{-1}$ and 5000 $fb^{-1}$. We consider $\tan\beta=35$ and
$M_A=400$ $GeV$.

\begin{figure}[t]
\begin{center}
\includegraphics[width=12.8cm]{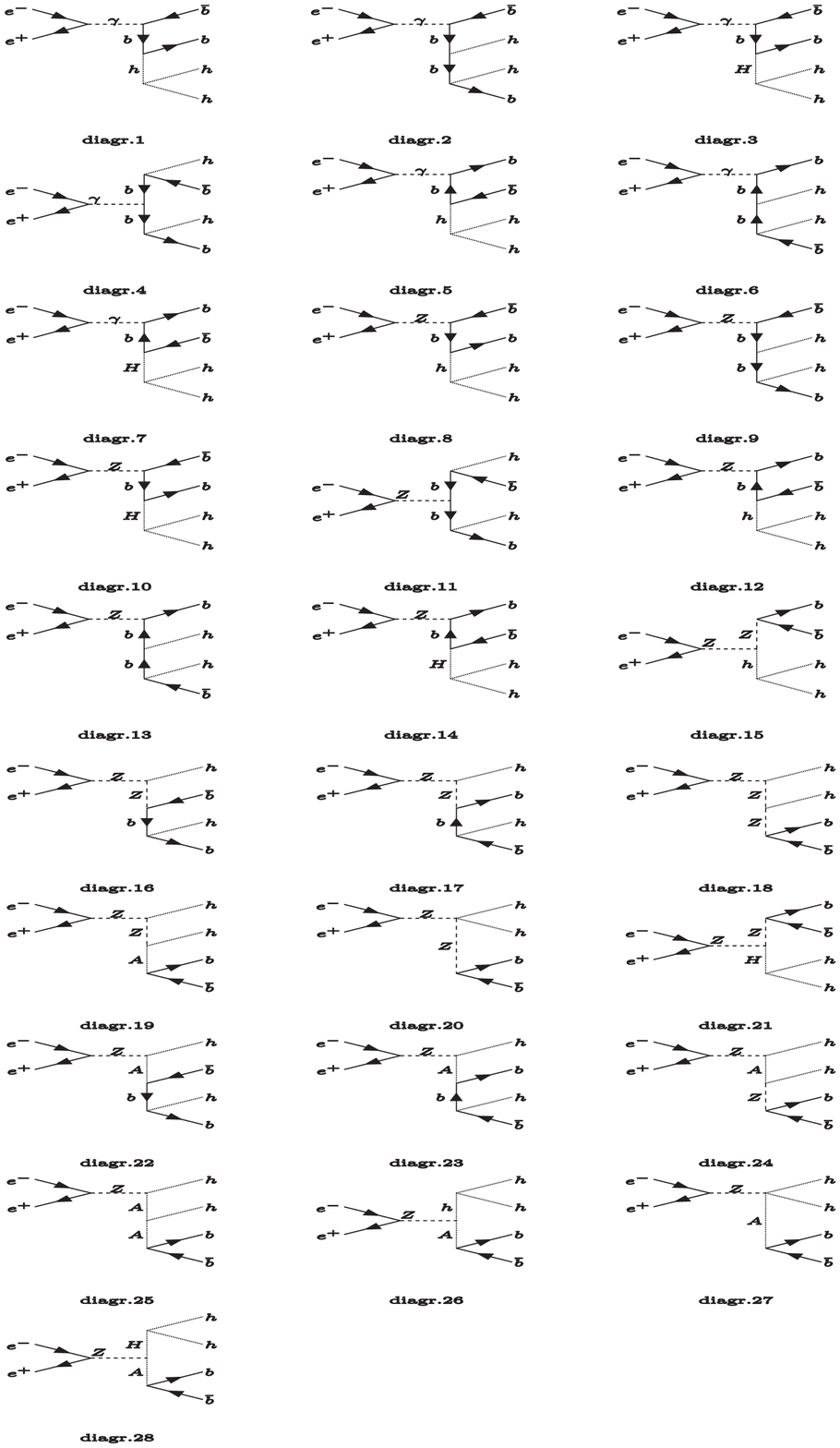}
\end{center}
\caption{Feynman diagrams at tree-level for $e^{+}e^{-}
\rightarrow b\bar b hh$. The diagrams for $e^{+}e^{-} \rightarrow
t\bar t hh$ are similar.} \label{f1}
\end{figure}

\begin{figure}[t]
\begin{center}
\includegraphics[width=12.2cm]{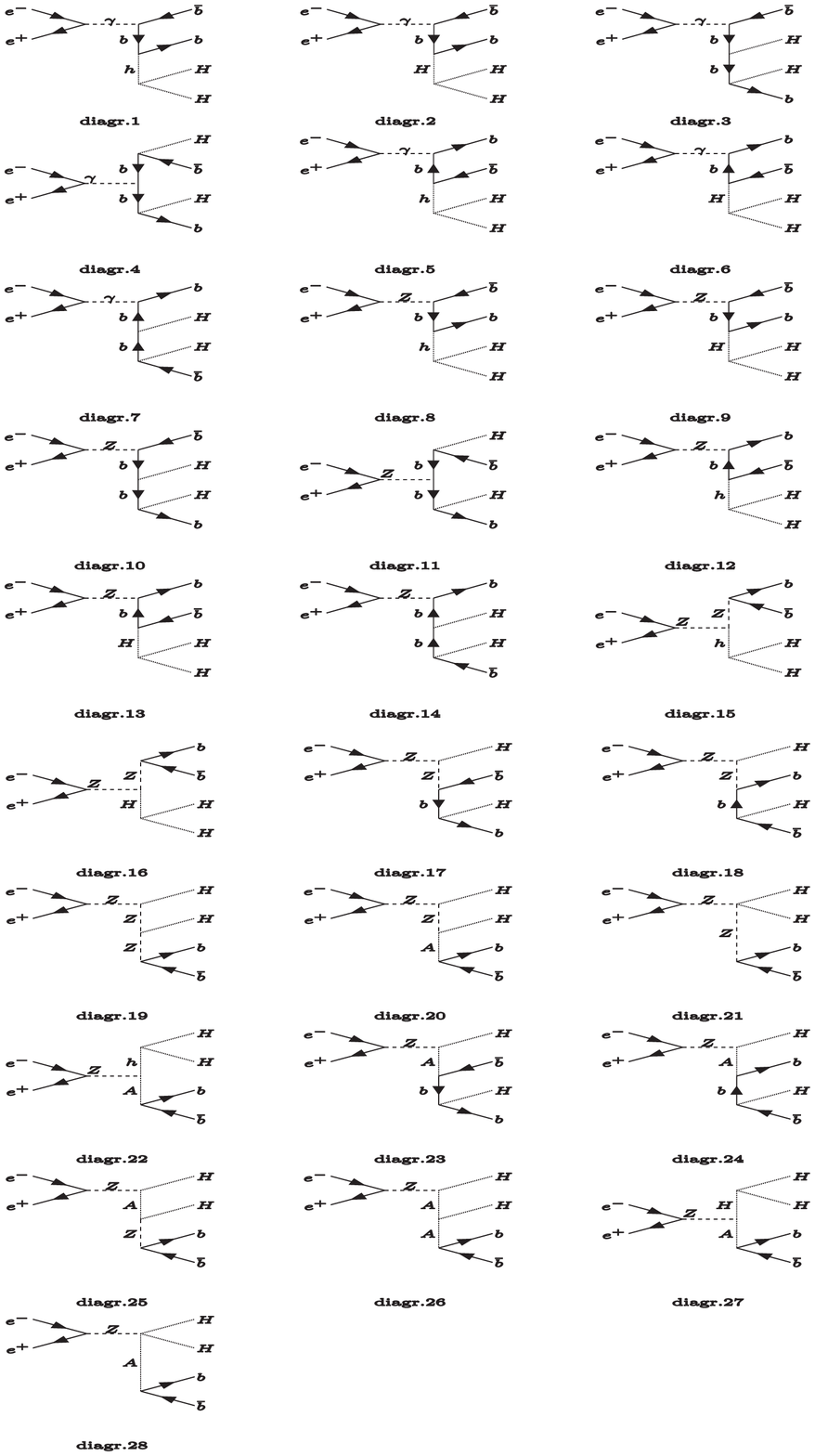}
\end{center}
\caption{Feynman diagrams at tree-level for $e^{+}e^{-}
\rightarrow b\bar b HH$. The diagrams for $e^{+}e^{-} \rightarrow
t\bar t HH$ are similar.}\label{f2}
\end{figure}

\begin{figure}[t]
\begin{center}
\includegraphics[width=12.5cm]{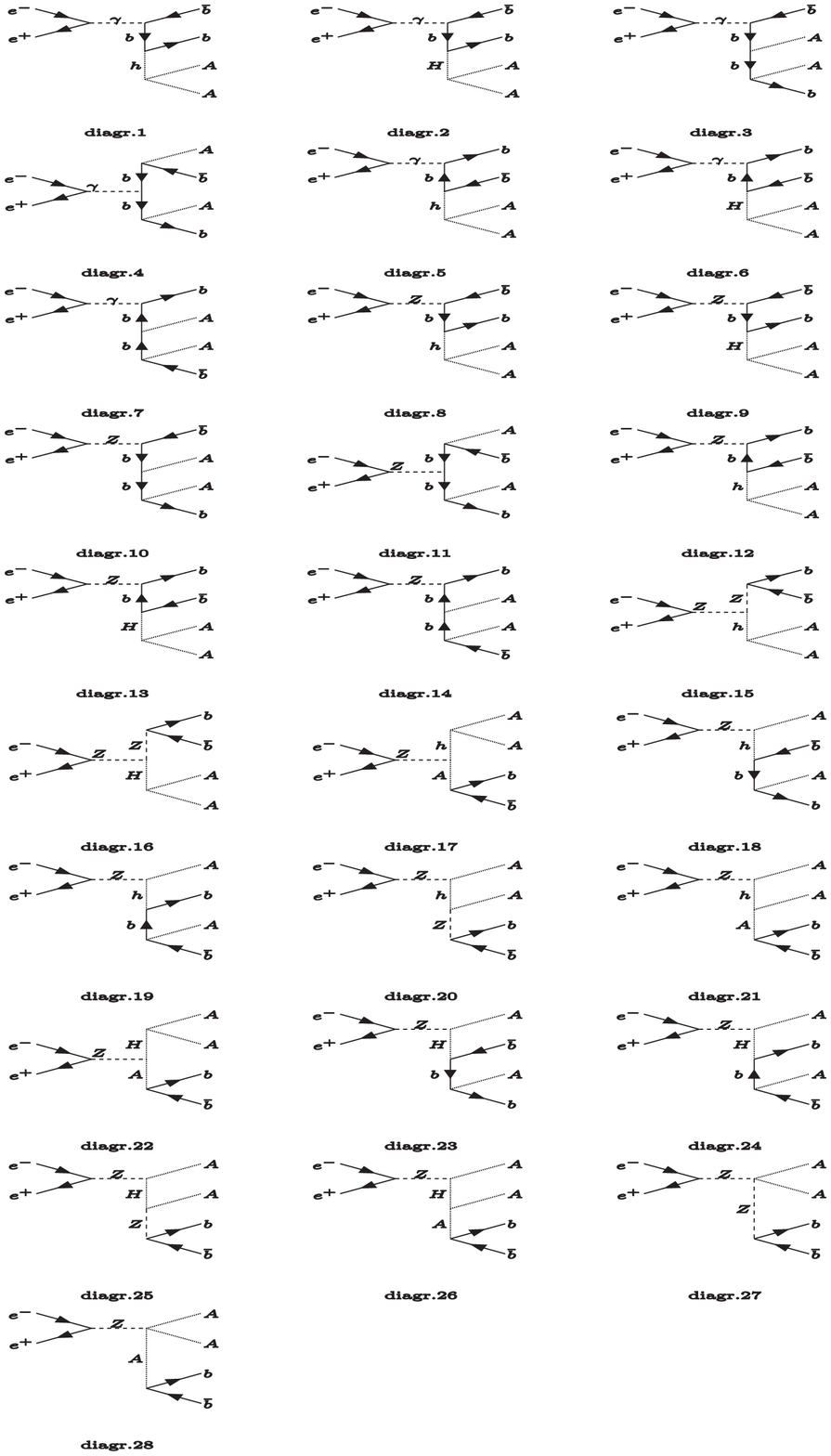}
\end{center}
\caption{Feynman diagrams at tree-level for $e^{+}e^{-}
\rightarrow b\bar b AA$. The diagrams for $e^{+}e^{-} \rightarrow
t\bar t AA$ are similar.}\label{f3}
\end{figure}

The Higgs couplings with quarks, the largest couplings in the
MSSM, are directly accessible in the processes where the Higgs
boson is radiated off bottom quarks, $e^{+}e^{-}\rightarrow b\bar
b hh, b\bar bHH, b\bar b AA$ as well as in the processes
$e^{+}e^{-}\rightarrow t\bar t hh, t\bar t HH, t\bar t AA$ where
the Higgs boson is radiated off top quarks. These processes depend
on the Higgs boson triple self-couplings, which could lead us to
obtain the first non-trivial information on the Higgs potential.
We are interested in finding regions that could allow the
observation of the $b\bar b hh, b\bar bHH, b\bar b AA$ and $t\bar
t hh, t\bar tHH, t\bar t AA$ processes at future linear
$e^{+}e^{-}$ colliders energies: ILC and CLIC. We consider the
complete set of Feynman diagrams at tree-level
(Figs.~\ref{f1}-\ref{f3}) and use the CALCHEP \cite{Pukhov}
packages to evaluate the amplitudes and cross-sections of the
processes $e^{+}e^{-}\rightarrow b \bar b h_ih_i, t \bar t
h_ih_i$.

This paper is organized as follows: In Sec. II, the Higgs boson
interaction Lagrangian and the self-couplings are presented. In
Sec. III, we study the triple Higgs boson self-coupling through
the processes $e^{+}e^{-}\rightarrow b \bar b hh (HH, AA)$ and
$e^{+}e^{-}\rightarrow t \bar t hh (HH, AA)$ at future
$e^{+}e^{-}$ colliders energies (ILC/CLIC) and, finally, we
summarize our results in Sec. IV.

\section{Higgs Boson Interaction Lagrangian}

The Higgs sector of MSSM includes five physical fields: two
neutral CP-even Higgs scalar $(h^0, H^0)$, one neutral CP-odd
Higgs scalar $(A^0)$ and a charged Higgs pair $(H^\pm)$. The Higgs
boson interaction lagrangian has the form

\begin{equation}
{\cal L}^{Higgs}_{Int}={\cal L}^{(3)}_{Int}+{\cal L}^{(4)}_{Int},
\end{equation}

\noindent where ${\cal L}^{(3)}_{Int}$ is the lagrangian of the
triple Higgs boson interactions, while ${\cal L}^{(4)}_{Int}$ is
the lagrangian of the quartic Higgs boson interactions.
Explicitly, ${\cal L}^{(3)}_{Int}$ is:

\begin{eqnarray}
{\cal
L}^{(3)}_{Int}&=&\frac{\lambda_{hhh}}{3!}hhh+\frac{\lambda_{hhH}}{2!}hhH+
\frac{\lambda_{hHH}}{2!}hHH+\frac{\lambda_{HHH}}{3!}HHH+\frac{\lambda_{hAA}}{2!}hAA \nonumber\\
&+&\frac{\lambda_{HAA}}{2!}HAA+
\lambda_{hH^+H^-}hH^+H^-+\lambda_{HH^+H^-}HH^+H^-.
\end{eqnarray}

The triple Higgs self-coupling of the MSSM can be parameterized as
\cite{Li,Gunion2,Gunion3,Gunion4,Okada,Haber1,Ellis,Dress,Djouadi2,Djouadi3,Osland0,Boudjema,Djouadi4,Ferrera,Osland,Osland2,Osland1,Abdesslam,Lafaye,Miller,Vernon},

\begin{eqnarray}
\lambda_{hhh}&=&3\cos2\alpha
\sin(\beta+\alpha)+3\frac{\epsilon}{M^2_Z}\frac{\cos\alpha}{\sin\beta}\cos^2\alpha,\nonumber\\
\lambda_{Hhh}&=&2\sin2\alpha \sin(\beta+\alpha)-\cos2\alpha
\cos(\beta+\alpha)+3\frac{\epsilon}{M^2_Z}\frac{\sin\alpha}{\sin\beta}\cos^2\alpha,\nonumber\\
\lambda_{HHh}&=&-2\sin2\alpha \cos(\beta+\alpha)-\cos2\alpha
\sin(\beta+\alpha)+3\frac{\epsilon}{M^2_Z}\frac{\cos\alpha}{\sin\beta}\sin^2\alpha,\nonumber\\
\lambda_{HHH}&=&3\cos2\alpha
\cos(\beta+\alpha)+3\frac{\epsilon}{M^2_Z}\frac{\sin\alpha}{\sin\beta}\sin^2\alpha,\\
\lambda_{hAA}&=&\cos2\beta
\sin(\beta+\alpha)+\frac{\epsilon}{M^2_Z}\frac{\cos\alpha}{\sin\beta}\cos^2\beta,\nonumber\\
\lambda_{HAA}&=&-\cos2\beta
\cos(\beta+\alpha)+\frac{\epsilon}{M^2_Z}\frac{\sin\alpha}{\sin\beta}\cos^2\beta,\nonumber
\end{eqnarray}

\noindent where  $\epsilon\approx
\frac{3G_Fm^4_t}{\sqrt{2}\pi^2\sin^2\beta}\log\frac{M^2_S}{m^2_t}$
and the connection between the mixing angles $\alpha$ and $\beta$
is given by

\begin{equation}
\tan2\alpha=\tan2\beta
\frac{M^2_A+M^2_Z}{M^2_A-M^2_Z+\epsilon/\cos\beta} \hspace{5mm}
\mbox{with} \hspace{5mm} -\frac{\pi}{2}\leq \alpha \leq 0,
\end{equation}

\noindent as a function of $M_A$ and $\tan\beta$.

\section{Higgs Bosons Pair Production with Modified Triple
Self-Coupling}

In this section we present numerical results for
$e^{+}e^{-}\rightarrow b \bar b hh(HH, AA)$ and
$e^{+}e^{-}\rightarrow t \bar t hh(HH, AA)$ with double Higgs
bosons production. We carry out the calculations using the
framework of the Minimal Supersymmetric Standard Model at the
future linear $e^{+}e^{-}$ colliders. We use the CALCHEP
\cite{Pukhov} packages for calculations of the matrix elements and
cross-sections. These packages provide automatic computation of
the cross-sections and distributions in the MSSM as well as their
extensions at tree-level. We consider the high energy stage of a
possible Future Linear $e^{+}e^{-}$ Collider (ILC) with
$\sqrt{s}=500, 1000, 1600$ $GeV$ and design luminosity 1000
$fb^{-1}$. For the numerical computation, we have adopted the
following parameters: the angle of Weinber $\sin^2\theta_W=0.232$,
the mass ($m_b=4.5$ $GeV$) of the bottom quark, the mass
($m_t=175$ $GeV$) of the top quark, the mass ($m_{Z^0}=91.2$
$GeV$) of the $Z^0$, the mass ($M_A=200-400$ $GeV$) of the CP-odd
Higgs scalar and $\tan\beta=35$.

\subsection{Triple Higgs Bosons Self-Coupling Via $e^{+}e^{-}\rightarrow b \bar b hh(HH,AA)$}

\begin{figure}[t]
\begin{center}
\includegraphics[width=11cm]{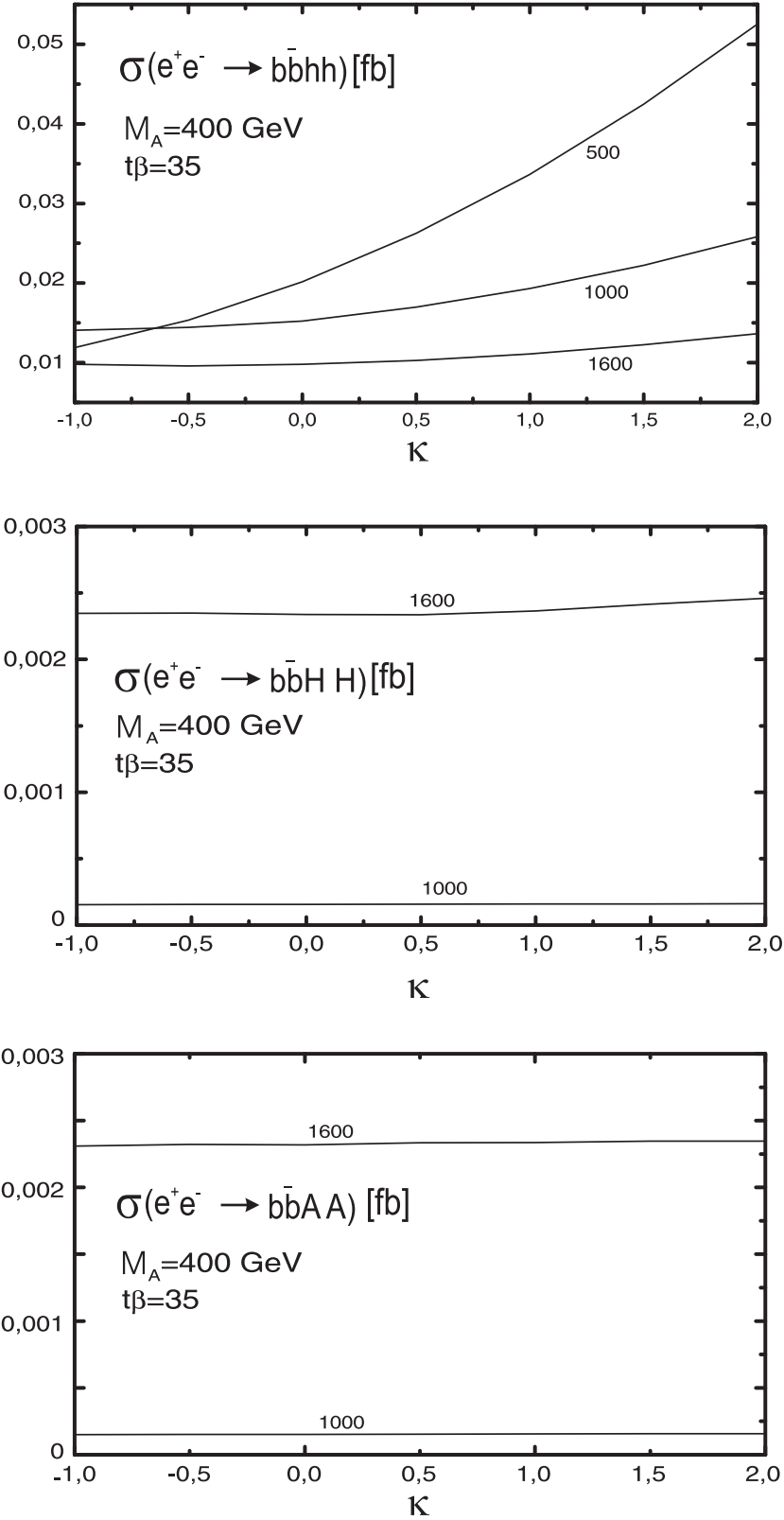}
\end{center}
\caption{Variation of the cross-section $\sigma(b\bar b hh, b\bar
b HH,b\bar b AA)$ with the modified triple coupling
$\kappa\lambda_{hhh}$, $\kappa\lambda_{Hhh}$ at a collider energy
of $\sqrt{s}= 500, 1000, 1600$ $GeV$ with $M_A= 400$ $GeV$ and
$\tan\beta=35$.}\label{f4}
\end{figure}

\begin{figure}[t]
\begin{center}
\includegraphics[width=11cm]{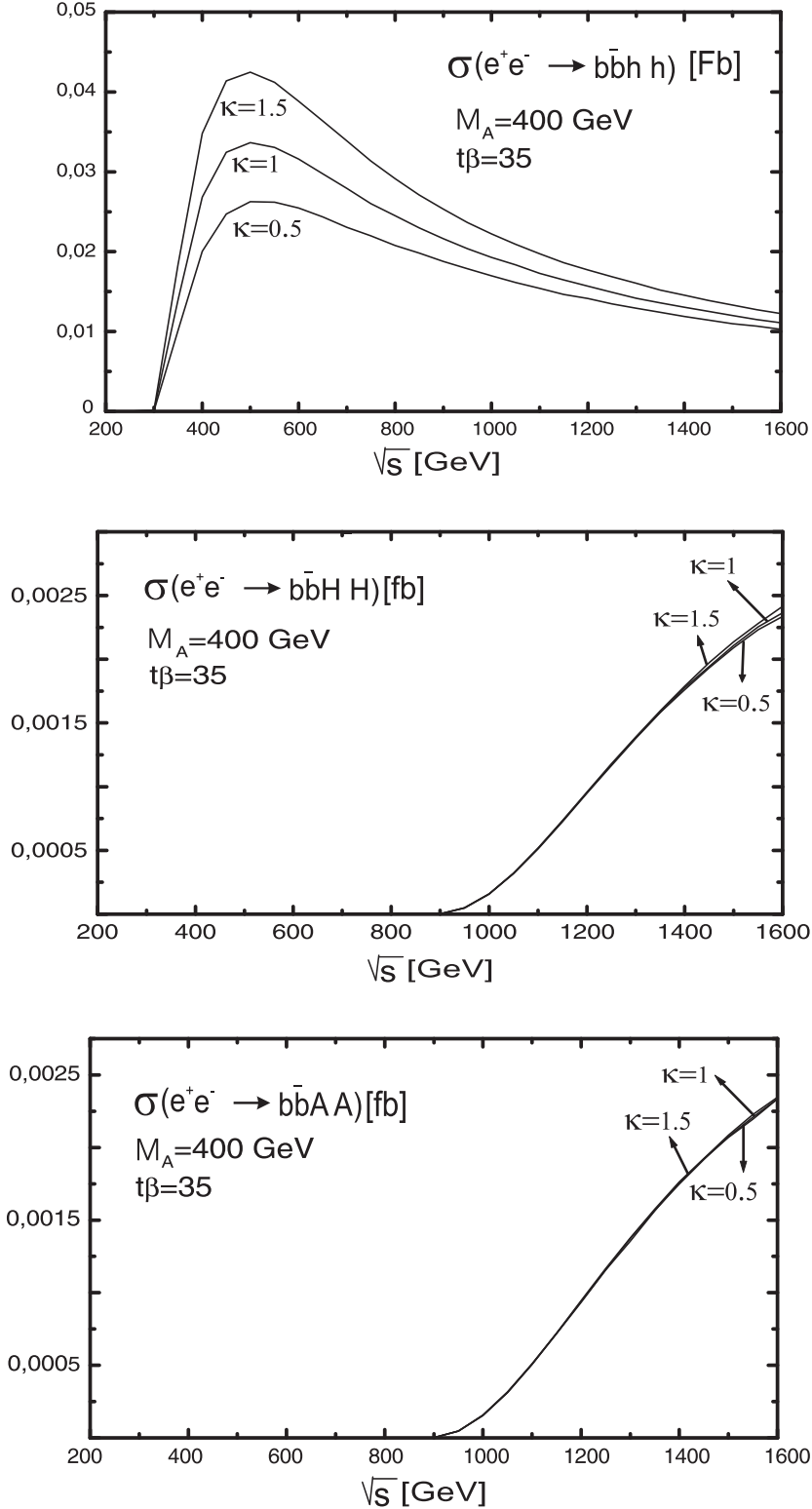}
\end{center}
\caption{The dependence of the cross-section on center-of-mass
energy $\sqrt{s}$ for $M_A= 400$ $GeV$ and $\tan\beta=35$. The
variation of the cross-section for modified triple couplings
$\kappa \lambda_{hhh}$, $\kappa \lambda_{Hhh}$ is indicated by
$\kappa=0.5, 1.5$.}\label{f5}
\end{figure}

To illustrate our results on the sensitivity to the $hhh, Hhh,
HHh, HHH, hAA, HAA$ triple modified Higgs bosons self-coupling, we
show the $\kappa$ dependence of the total cross-section for
$e^{+}e^{-}\rightarrow b \bar b hh(HH, AA)$ in Fig.~\ref{f4}. We
consider one representative value of the Higgs boson mass,
$M_A=400$ $GeV$, and $\tan\beta=35$ with a center-of-mass energy
of $\sqrt{s}=500, 1000, 1600$ GeV and vary the triple couplings
$\kappa \lambda_{hhh}, \kappa \lambda_{Hhh}, \kappa \lambda_{HHh},
\kappa \lambda_{HHH}, \kappa \lambda_{hAA}, \kappa \lambda_{HAA}$
within the range $\kappa=-1$ and $+2$. In all cases, the
cross-section is sensitive to the value of the triple couplings,
as well as in the case of the process $e^{+}e^{-}\rightarrow b
\bar b hh$, where for large values of $M_A$ (the decoupling
limit), the corresponding MSSM triple coupling approaches the SM
triple coupling \cite{A.Gutierrez6}. However, the cross-section
and its sensibility to $ \lambda_{hhh}, \lambda_{hHH}$ decrease
with increasing collider energy for $e^{+}e^{-}\rightarrow b \bar
b hh$, while for $e^{+}e^{-}\rightarrow t \bar t HH(AA)$, the
sensitivity to $ \lambda_{hHH}, \lambda_{HHH}$ and $
\lambda_{hAA}, \lambda_{HAA}$ increases with rising collider
energy. In particular, we can see in Fig.~\ref{f5} for
$e^{+}e^{-}\rightarrow b \bar b hh(HH, AA)$, the maximum
cross-section (for $M_A=400$ $GeV$ and $\tan\beta=35$) is reached
for $\sqrt{s}\sim 500$ $GeV$ and $1800$ $GeV$ respectively, which
is consistent with Ref. \cite{A.Gutierrez6} in the case where $h$
is identified as the Higgs boson of the standard model. As an
indicator of the order of magnitude, in Tables~\ref{t1}-\ref{t3}
we present the Higgs boson number of events (we have to
multiplicate by the corresponding Branching Ratios to obtain the
observable number of events) for several Higgs boson masses $M_A$,
center-of-mass energy and $\kappa$ values and for a luminosity of
$1000 fb^{-1}$. If we consider the $h\rightarrow b \bar b$ decay
for $M_h < 130$ $GeV$, we have some opportunity to detect this
process. In this region, the number of events is small but
sufficient to detect $e^{+}e^{-}\rightarrow b \bar b hh\rightarrow
b \bar b b \bar b b \bar b$. The $BR(h\rightarrow b \bar b)\sim
0.6$ and the background for 6 b-jet are small.

Finally, for completeness, in Fig.~\ref{f6} we include a contour
plot for the number of events of the studied processes as a
function of $\sqrt{s}$ and $\kappa$ with $M_A= 400$ $GeV$ and
$\tan\beta=35$. These contour are obtained from
Tables~\ref{t1}-\ref{t3}. Because in major parts of the MSSM
parameter space the neutral Higgs bosons $H$ and $A$ are quite
heavy, it is difficult to detect the processes
$e^{+}e^{-}\rightarrow b \bar b HH(AA)$ when the relevant
mechanism is $e^{+}e^{-}\rightarrow b \bar b hh$, as shown in the
number of events given in Tables~\ref{t1}-\ref{t3}.

\vspace{5mm}

\begin{table}[t]
\caption{Total production of Higgs boson pairs in the MSSM for
$\tan\beta=35$, ${\cal L}=1000$ $fb^{-1}$ and $\kappa=0.5$.}
\label{t1}
\begin{center}
\begin{tabular}{c|c|c|c}
\hline
Total Production of Higgs Boson Pairs & \multicolumn{3}{c}{$e^{+}e^{-}\rightarrow b \bar b hh(HH, AA) \hspace{8mm} \kappa=0.5$}\\
\hline \hline
\cline{2-4} & $\sqrt{s}= $ & $\sqrt{s}= $ & $\sqrt{s}= $  \\
$M_A(GeV)$ & 500 $GeV$ & 1000 $GeV$ & 1600 $GeV$ \\
\hline \hline
 200 & 26 (-,-)  & 17 (7,7) & 10 (7,7) \\
 250 & 26 (-,-)  & 17 (7,7) & 10 (7,7) \\
 300 & 26 (-,-)  & 17 (7,7) & 10 (7,7)  \\
 350 & 26 (-,-)  & 17 (7,7) & 10 (7,7)  \\
 400 & 26 (-,-)  & 17 (7,7) & 10 (7,7)  \\
\hline
\end{tabular}
\end{center}
\end{table}

\begin{table}[t]
\caption{Total production of Higgs boson pairs in the MSSM for
$\tan\beta=35$, ${\cal L}=1000$ $fb^{-1}$ and $\kappa=1(MSSM)$. }
\label{t2}
\begin{center}
\begin{tabular}{c|c|c|c}
\hline
Total Production of Higgs Boson Pairs & \multicolumn{3}{c}{$e^{+}e^{-}\rightarrow b \bar b hh(HH, AA) \hspace{8mm} \kappa=1(MSSM)$}\\
\hline \hline
\cline{2-4} & $\sqrt{s}= $ & $\sqrt{s}= $ & $\sqrt{s}= $  \\
$M_A(GeV)$ & 500 $GeV$ & 1000 $GeV$ & 1600 $GeV$ \\
\hline \hline
 200 & 33 (-,-) & 19 (7,7) & 11 (7,7) \\
 250 & 33 (-,-) & 19 (7,7) & 11 (7,7) \\
 300 & 33 (-,-) & 19 (7,7) & 11 (7,7) \\
 350 & 33 (-,-) & 19 (7,7) & 11 (7,7)  \\
 400 & 33 (-,-) & 19 (7,7) & 11 (7,7)  \\
\hline
\end{tabular}
\end{center}
\end{table}

\begin{table}[t]
\caption{Total production of Higgs boson pairs in the MSSM for
$\tan\beta=35$, ${\cal L}=1000$ $fb^{-1}$ and $\kappa=1.5$. }
\label{t3}
\begin{center}
\begin{tabular}{c|c|c|c}
\hline
Total Production of Higgs Boson Pairs & \multicolumn{3}{c}{$e^{+}e^{-}\rightarrow b \bar b hh(HH, AA) \hspace{8mm} \kappa=1.5$}\\
\hline \hline
\cline{2-4} & $\sqrt{s}= $ & $\sqrt{s}= $ & $\sqrt{s}= $  \\
$M_A(GeV)$ & 500 $GeV$ & 1000 $GeV$ & 1600 $GeV$ \\
\hline \hline
 200 & 42 (-,-) & 22 (7,7) & 12 (7,7) \\
 250 & 42 (-,-) & 22 (7,7) & 12 (7,7) \\
 300 & 42 (-,-) & 22 (7,7) & 12 (7,7) \\
 350 & 42 (-,-) & 22 (7,7) & 12 (7,7) \\
 400 & 42 (-,-) & 22 (7,7) & 12 (7,7)  \\
\hline
\end{tabular}
\end{center}
\end{table}

\subsection{Triple Higgs Bosons Self-Coupling Via $e^{+}e^{-}\rightarrow t \bar t hh(HH,AA)$}

\begin{figure}[t]
\begin{center}
\includegraphics[width=11cm]{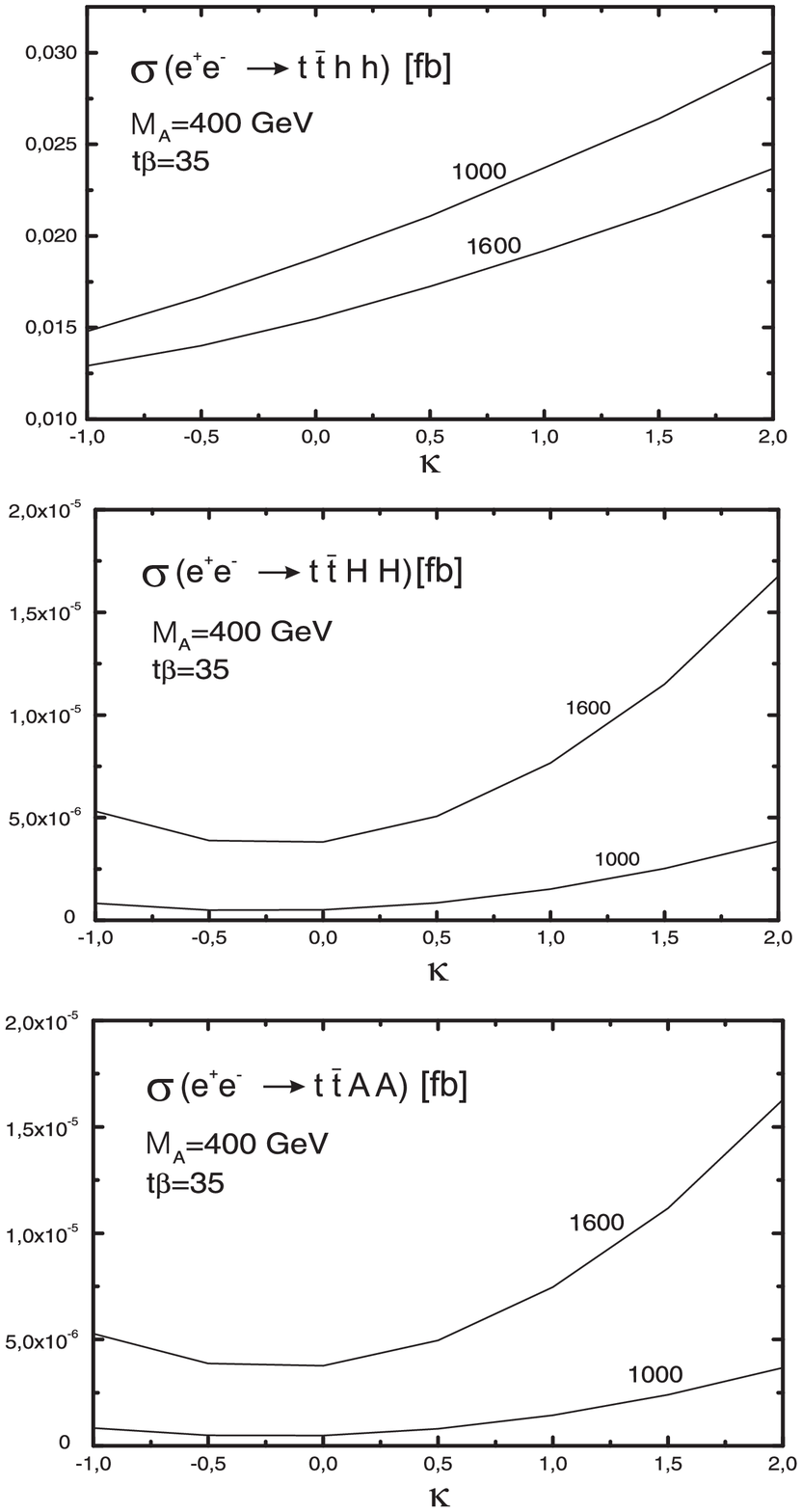}
\end{center}
\caption{The same as in Fig. 4, but for the processes $e^{+}e^{-}
\rightarrow t\bar t hh, t\bar t HH, t\bar t AA$.}\label{f7}
\end{figure}

\begin{figure}[t]
\begin{center}
\includegraphics[width=11cm]{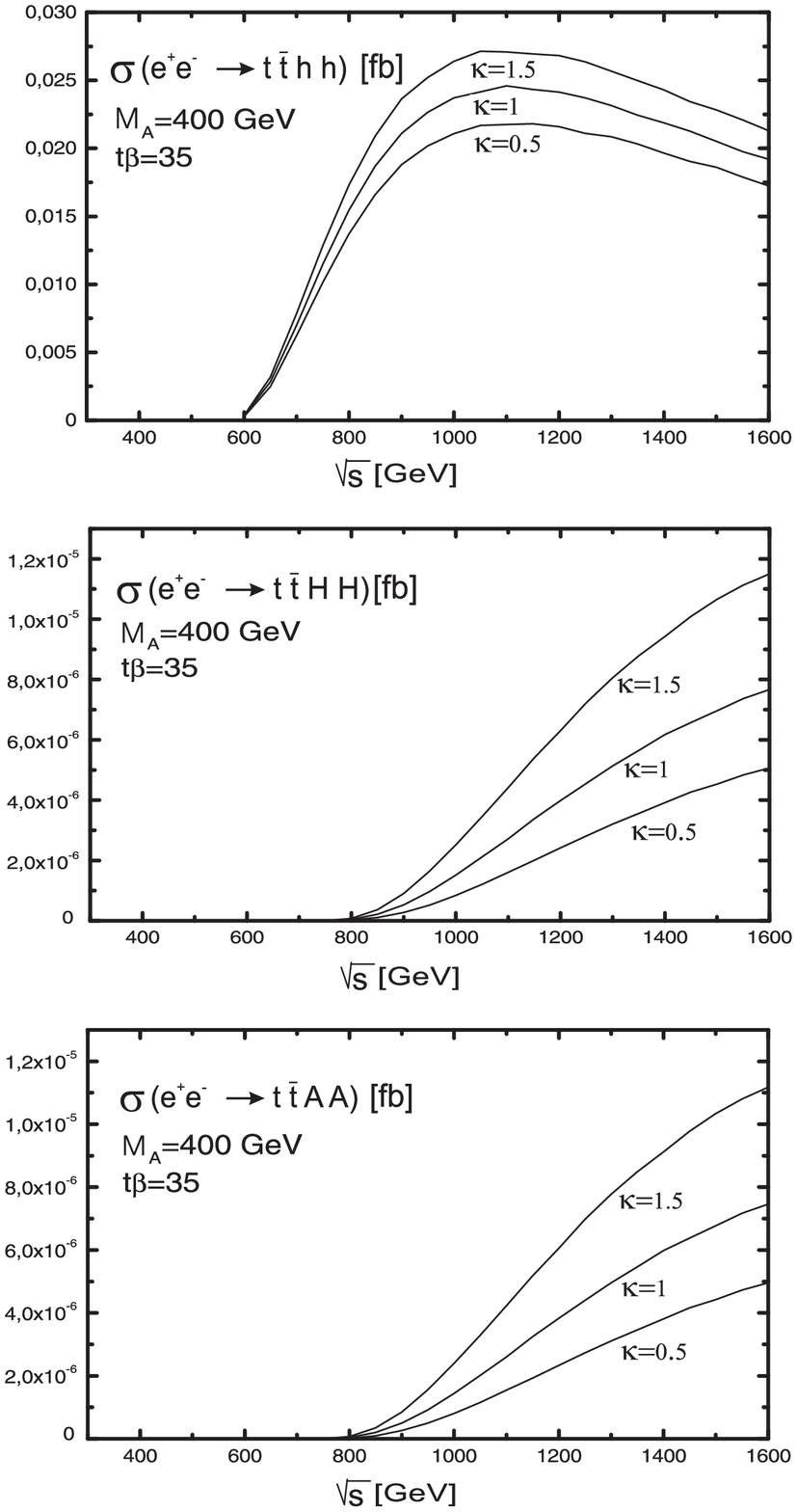}
\end{center}
\caption{The same as in Fig. 5, but for the processes $e^{+}e^{-}
\rightarrow t\bar t hh, t\bar t HH, t\bar t AA$.}\label{f8}
\end{figure}

As in the case of the processes $e^{+}e^{-}\rightarrow b \bar b
hh(HH,AA)$, Fig.~\ref{f7} shows the variation of the
cross-sections for $e^{+}e^{-}\rightarrow t \bar t hh(HH,AA)$ with
$\kappa\lambda_{hhh}, \kappa\lambda_{Hhh}, \kappa\lambda_{HHh},
\kappa\lambda_{HHH}, \kappa\lambda_{hAA}, \kappa\lambda_{HAA}$,
$\kappa=-1$ to $+2$, $500 \leq \sqrt{s}\leq 1600$ $GeV$, $M_A=
400$ $GeV$ and $\tan\beta=35$. The cross-section and its
sensibility to the triple self-coupling on the energy range
considered is dependent on the behavior of the cross-section with
the center-of-mass energy determined by the position of the
maximum. In Fig.~\ref{f8}, we can see that the maximum
cross-section is reached for $\sqrt{s}\approx 1100$ $GeV$ in the
case of $e^{+}e^{-}\rightarrow t \bar t hh$, while for
$e^{+}e^{-}\rightarrow t \bar t HH(AA)$ it is reached for
$\sqrt{s}\approx 1800$ $GeV$. For the process
$e^{+}e^{-}\rightarrow t \bar t hh$ (see Fig.~\ref{f8}, Ref.
\cite{A.Gutierrez6}) and for large values of $M_A$ (the decoupling
limit), the corresponding MSSM triple coupling approaches the SM
triple coupling \cite{A.Gutierrez6}. On the other hand, for $t\bar
t HH(AA)$, the cross-section of the order $10^{-5}$ $fb$ is small
and is difficult to measure in the collider. In
Tables~\ref{t4}-\ref{t6} to indicate the order of magnitude, we
present the Higgs boson number of events (we have to multiplicate
for the corresponding Branching Ratios to obtain the observable
number of events) for several Higgs boson masses $M_A$,
center-of-mass energy and $\kappa$ values and for a luminosity of
$1000 fb^{-1}$.

For $e^{+}e^{-}\rightarrow t \bar t hh$, the most favorable
situation is for a center-of-mass energy of $1100$ GeV and $M_h <
130$ $GeV$, but in this case, the Branching Ratios for the four
decay modes make this process very much suppressed.

Finally, we include a contour plot for the number of events of the
studied processes in the ($\sqrt{s}, \kappa$) plane with $M_A=
400$ $GeV$ and $\tan\beta=35$ in Fig.~\ref{f9}. These contours are
obtained from Tables~\ref{t4}-\ref{t6}. Since in major parts of
the MSSM parameter space the neutral Higgs bosons $H$ and $A$ are
quite heavy, it is difficult to detect the processes
$e^{+}e^{-}\rightarrow t \bar t HH(AA)$ when the relevant
mechanism is $e^{+}e^{-}\rightarrow t \bar t hh$ for
$\sqrt{s}\approx 1100$ $GeV$ as shown in Tables~\ref{t4}-\ref{t6}.

\begin{table}[t]
\caption{Total production of Higgs boson pairs in the MSSM for
$\tan\beta=35$, ${\cal L}=1000$ $fb^{-1}$ and $\kappa=0.5$. }
\label{t4}
\begin{center}
\begin{tabular}{c|c|c|c}
\hline
Total Production of Higgs Boson Pairs & \multicolumn{3}{c}{$e^{+}e^{-}\rightarrow t \bar t hh(HH, AA) \hspace{8mm} \kappa=0.5$}\\
\hline \hline
\cline{2-4} & $\sqrt{s}= $ & $\sqrt{s}= $ & $\sqrt{s}= $  \\
$M_A(GeV)$ & 500 $GeV$ & 1000 $GeV$ & 1600 $GeV$ \\
\hline \hline
 200 & - (-,-) & 21 (-,-)& 17 (-,-)\\
 250 & - (-,-) & 21 (-,-)& 17 (-,-)\\
 300 & - (-,-) & 21 (-,-)& 17 (-,-) \\
 350 & - (-,-) & 21 (-,-)& 17 (-,-) \\
 400 & - (-,-) & 21 (-,-)& 17 (-,-) \\
\hline
\end{tabular}
\end{center}
\end{table}

\begin{table}[t]
\caption{Total production of Higgs boson pairs in the MSSM for
$\tan\beta=35$, ${\cal L}=1000$ $fb^{-1}$ and $\kappa=1(MSSM)$. }
\label{t5}
\begin{center}
\begin{tabular}{c|c|c|c}
\hline
Total Production of Higgs Boson Pairs & \multicolumn{3}{c}{$e^{+}e^{-}\rightarrow t \bar t hh (HH, AA) \hspace{8mm} \kappa=1(MSSM)$}\\
\hline \hline
\cline{2-4} & $\sqrt{s}= $ & $\sqrt{s}= $ & $\sqrt{s}= $  \\
$M_A(GeV)$ & 500 $GeV$ & 1000 $GeV$ & 1600 $GeV$ \\
\hline \hline
 200 & - (-,-) & 23 (-,-) & 19 (-,-) \\
 250 & - (-,-) & 23 (-,-) & 19 (-,-) \\
 300 & - (-,-) & 23 (-,-) & 19 (-,-) \\
 350 & - (-,-) & 23 (-,-) & 19 (-,-)  \\
 400 & - (-,-) & 23 (-,-) & 19 (-,-)  \\
\hline
\end{tabular}
\end{center}
\end{table}

\begin{table}[t]
\caption{Total production of Higgs boson pairs in the MSSM for
$\tan\beta=35$, ${\cal L}=1000$ $fb^{-1}$ and $\kappa=1.5$. }
\label{t6}
\begin{center}
\begin{tabular}{c|c|c|c}
\hline
Total Production of Higgs Boson Pairs & \multicolumn{3}{c}{$e^{+}e^{-}\rightarrow t \bar t hh (HH, AA) \hspace{8mm} \kappa=1.5$}\\
\hline \hline
\cline{2-4} & $\sqrt{s}= $ & $\sqrt{s}= $ & $\sqrt{s}= $  \\
$M_A(GeV)$ & 500 $GeV$ & 1000 $GeV$ & 1600 $GeV$ \\
\hline \hline
 200 & - (-,-) & 26 (-,-) & 21 (-,-) \\
 250 & - (-,-) & 26 (-,-) & 21 (-,-) \\
 300 & - (-,-) & 26 (-,-) & 21 (-,-) \\
 350 & - (-,-) & 26 (-,-) & 21 (-,-) \\
 400 & - (-,-) & 26 (-,-) & 21 (-,-)  \\
\hline
\end{tabular}
\end{center}
\end{table}

\subsection{Triple Higgs Boson Self-Coupling Via $e^{+}e^{-}\rightarrow b \bar b  hh(HH,AA), t \bar t hh(HH,AA)$
at CLIC energies}

\begin{figure}[t]
\begin{center}
\includegraphics[width=11cm]{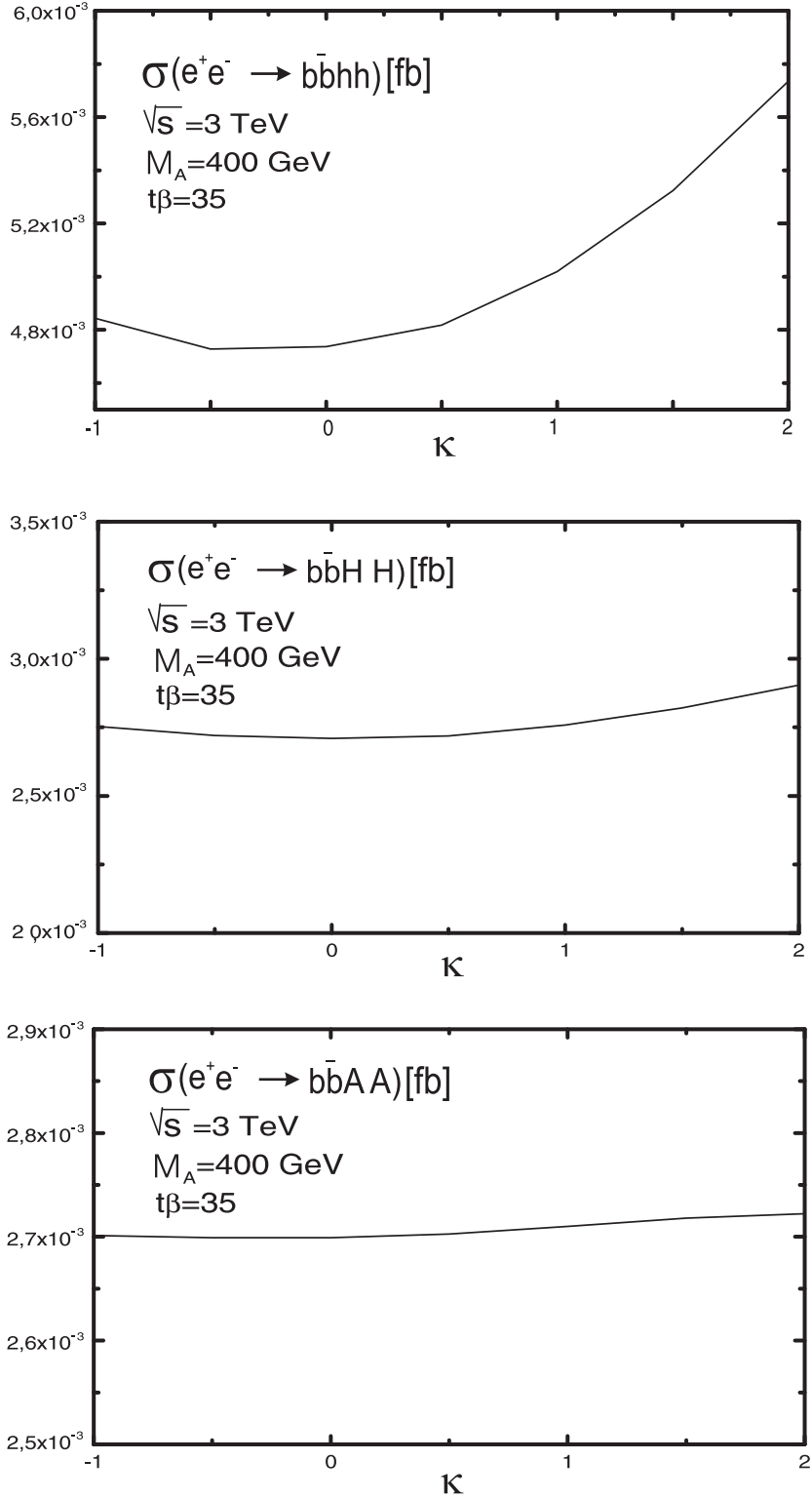}
\end{center}
\caption{The same as in Fig. 4, but for $\sqrt{s}= 3$ $TeV$ with
$M_A= 400$ $GeV$ and $\tan\beta=35$.}\label{f10}
\end{figure}

\begin{figure}[t]
\begin{center}
\includegraphics[width=13cm]{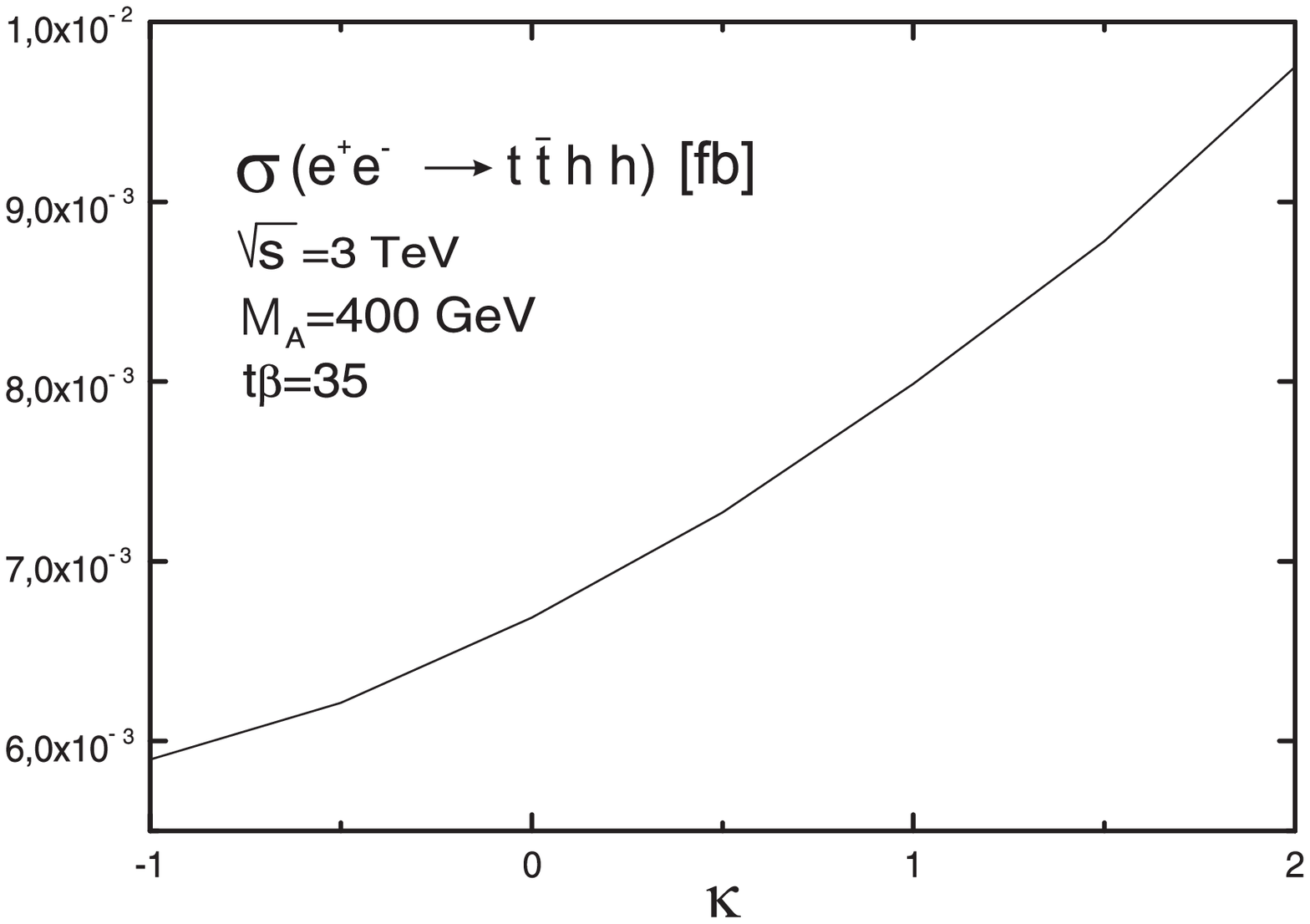}
\end{center}
\caption{The same as in Fig. 10, but for the process $e^{+}e^{-}
\rightarrow t\bar t hh$.}\label{f11}
\end{figure}

In this subsection we analyze the triple Higgs self-coupling
$\kappa \lambda_{hhh}, \kappa \lambda_{Hhh}, \kappa \lambda_{HHh},
\kappa \lambda_{HHH}, \kappa \lambda_{hAA}, \kappa \lambda_{HAA}$
via the processes $e^{+}e^{-}\rightarrow b \bar b  hh(HH,AA)$ and
$e^{+}e^{-}\rightarrow t \bar t hh(HH,AA)$ for energies expected
at the CLIC \cite{Accomando}. Figures~\ref{f10} and ~\ref{f11}
show the total cross-section for the double Higgs-strahlung in
$e^{+}e^{-}$ collisions, $e^{+}e^{-}\rightarrow b \bar b
hh(HH,AA)$ and $e^{+}e^{-}\rightarrow t \bar t hh$ as a function
of $\kappa$ for the c.m. energy of $\sqrt{s}=3$ $TeV$, $M_A=400$
$GeV$ and $\tan\beta=35$. From these figures, we observe that the
production cross-section of both processes is small, as it is of
the order of $10^{-3}$ $fb$.

Finally, in Tables~\ref{t7} and ~\ref{t8} we present the Higgs
boson ($hh, HH, AA$) number of events for several $\kappa$ values,
luminosities of 1000 and 5000 $fb^{-1}$ and center-of-mass energy
$\sqrt{s}=3$ $TeV$ (we have to multiplicate by the corresponding
Branching Ratios to obtain the observable number of events). It is
evident from Figures~\ref{f10} and ~\ref{f11} and Table~\ref{t7}
that it would be difficult to obtain a clear signal of the
processes $e^{+}e^{-}\rightarrow b \bar b hh(HH,AA)$,
$e^{+}e^{-}\rightarrow t \bar t hh(HH,AA)$ at energies of a future
linear collider such as CLIC, after having considered the
background, except for $\sqrt{s}=3$ $TeV$ and very high luminosity
$({\cal L}=5000$ $fb^{-1}$), as shown in Table~\ref{t8}. However,
for the CLIC center-of-mass energy, the $WW$ double Higgs fusion
process
\cite{Djouadi1,Djouadi2,Djouadi3,Osland0,Boudjema,Djouadi4}, which
increases with rising $\sqrt{s}$, can be exploited by large
energies and luminosities and would be the preferred mechanism to
measure the triple Higgs self-couplings.

\begin{table}[t]
\caption{Total production of Higgs boson pairs in the MSSM for
$\tan\beta=35$, $M_A=400$ $GeV$, $\sqrt{s}=3$ $TeV$ and ${\cal
L}=1000$ $fb^{-1}$.} \label{t7}
\begin{center}
\begin{tabular}{c|c|c|c|c}
\hline
Total Production of Higgs Boson Pairs & \multicolumn{4}{c}{$M_A=400$ $GeV$, \hspace{8mm} $\sqrt{s}=3$ $TeV$}\\
\hline \hline
\cline{2-5} & $e^{+}e^{-}\rightarrow$ & $e^{+}e^{-}\rightarrow$ & $e^{+}e^{-}\rightarrow$&  $e^{+}e^{-}\rightarrow$ \\
$\kappa$ & $b \bar b hh$ & $b \bar b HH$ & $b \bar b AA$ & $t \bar t hh$ \\
\hline \hline
0.5 & 5 & 3  & 3 & 7\\
1 &   5 & 3  & 3 & 8\\
1.5 & 5 & 3  & 3 & 9\\
\hline
\end{tabular}
\end{center}
\end{table}

\begin{table}[t]
\caption{Total production of Higgs boson pairs in the MSSM for
$\tan\beta=35$, $M_A=400$ $GeV$, $\sqrt{s}=3$ $TeV$ and ${\cal
L}=5000$ $fb^{-1}$.} \label{t8}
\begin{center}
\begin{tabular}{c|c|c|c|c}
\hline
Total Production of Higgs Boson Pairs & \multicolumn{4}{c}{$M_A=400$ $GeV$, \hspace{8mm} $\sqrt{s}=3$ $TeV$}\\
\hline \hline
\cline{2-5} & $e^{+}e^{-}\rightarrow$ & $e^{+}e^{-}\rightarrow$ & $e^{+}e^{-}\rightarrow$&  $e^{+}e^{-}\rightarrow$ \\
$\kappa$ & $b \bar b hh$ & $b \bar b HH$ & $b \bar b AA$ & $t \bar t hh$ \\
\hline \hline
0.5 & 25 & 15  & 14 & 35\\
1 &   26 & 16  & 15 & 41\\
1.5 & 28 & 18  & 17 & 47\\
\hline
\end{tabular}
\end{center}
\end{table}

\section{Conclusions}

$e^{+}e^{-}$ linear colliders represent a possible opportunity for
the triple Higgs boson self-coupling analysis. Therefore, we have
analyzed the neutral Higgs bosons self-couplings with a complete
set of the tree-level Feynman diagrams in the framework of the
MSSM. The dependence of the triple Higgs bosons self-coupling
$\lambda_{hhh}$, $\lambda_{Hhh}$, $\lambda_{hAA}$,
$\lambda_{HAA}$, $\lambda_{hHH}$ and $\lambda_{HHH}$ on
$\tan\beta$ and energy $\sqrt{s}$ were evaluated.

\begin{figure}[t]
\begin{center}
\includegraphics[width=11cm]{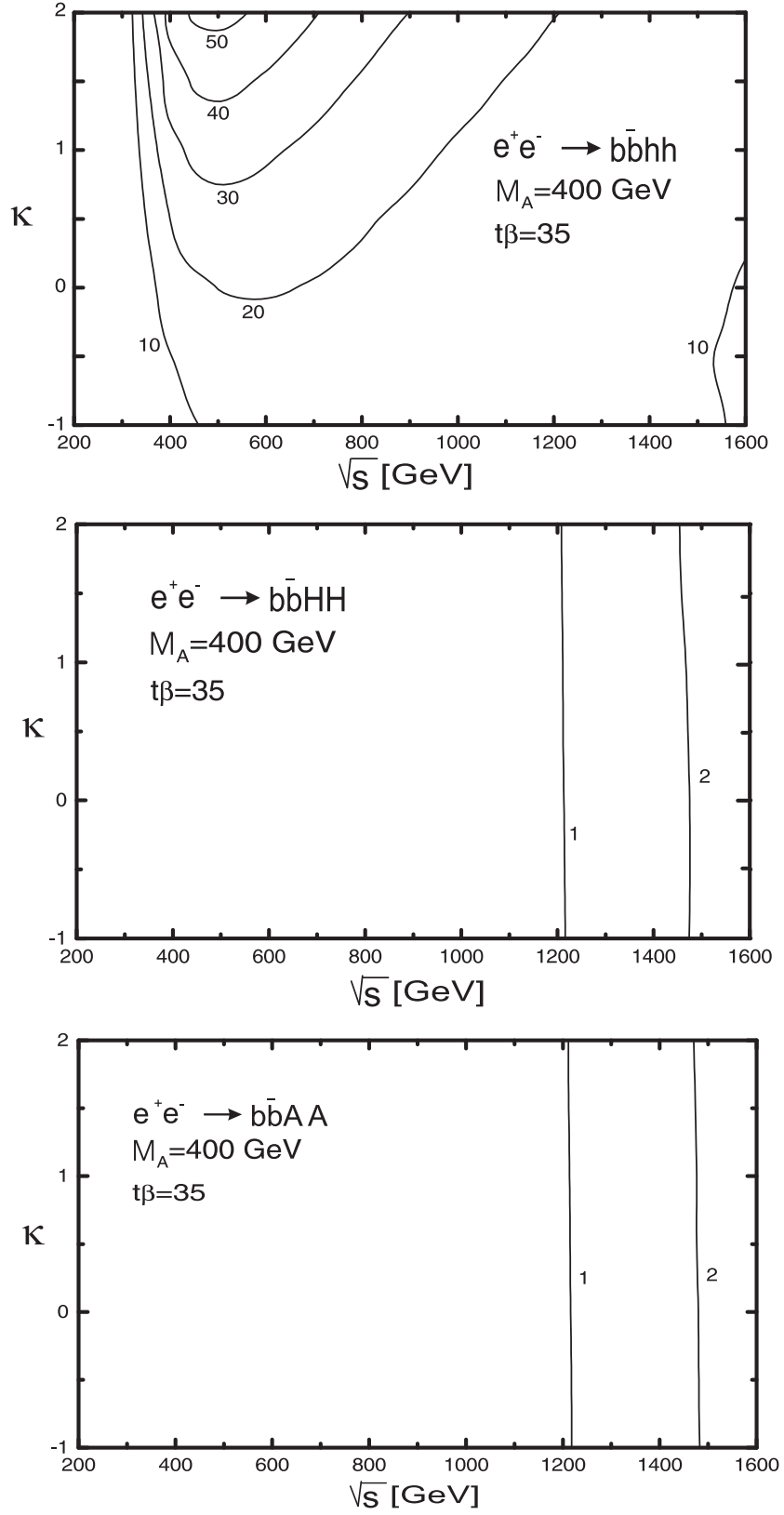}
\end{center}
\caption{Contour plot for the number of events of the processes
$e^{+}e^{-}\rightarrow b\bar b hh, b\bar b HH, b\bar b AA$ as a
function of $\sqrt{s}$ and $\kappa$.}\label{f6}
\end{figure}

\begin{figure}[t]
\begin{center}
\includegraphics[width=13cm]{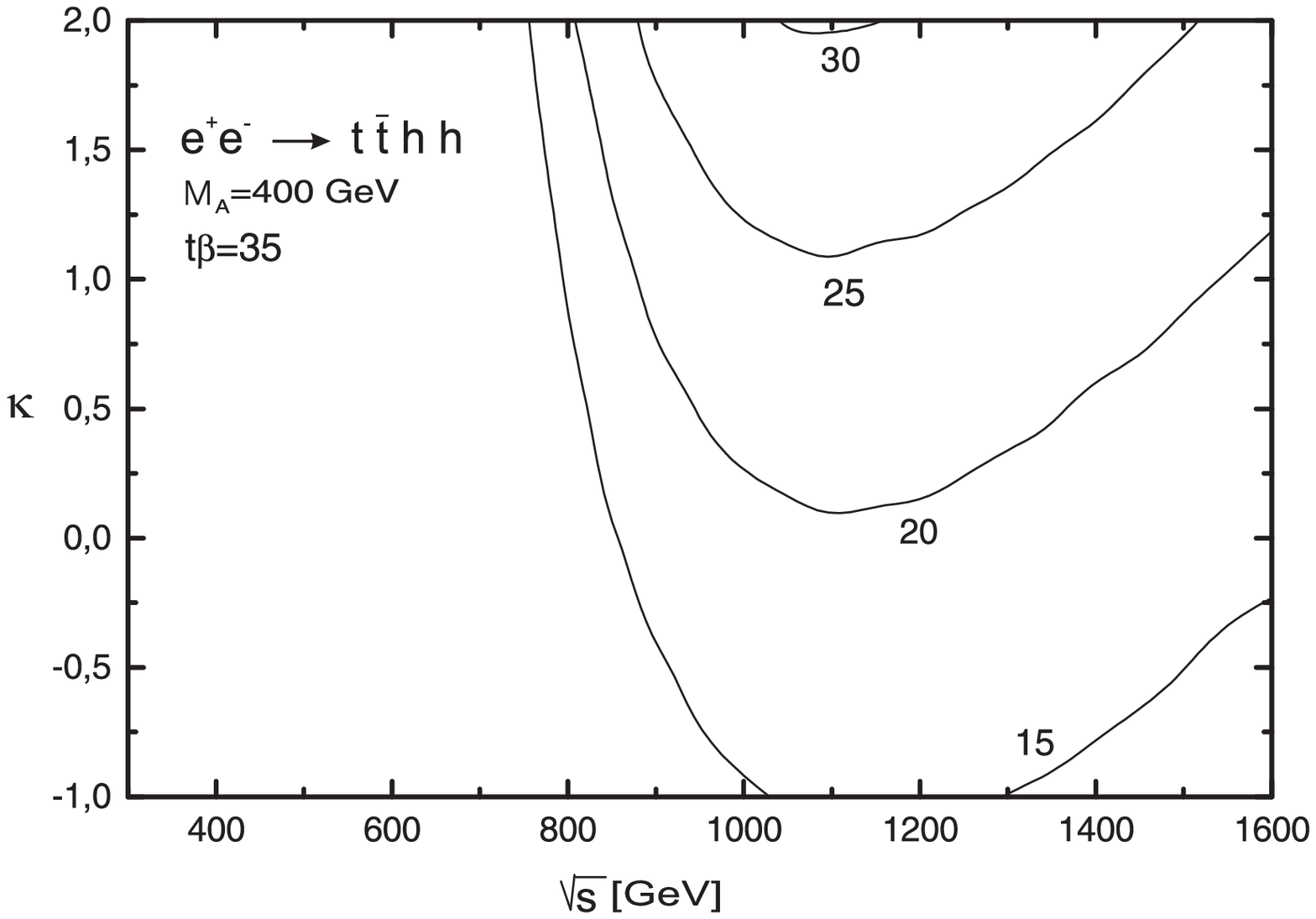}
\end{center}
\caption{The same as in Fig. 6, but for the process $e^{+}e^{-}
\rightarrow t\bar t hh$.}\label{f9}
\end{figure}

The extended Higgs spectrum in supersymmetric theories gives rise
to a plethora of triple couplings. The $hhh$ coupling is generally
quite different from the one expected in the standard model. It
can be measured in Higgs double production at Future International
Linear $e^{+}e^{-}$ Colliders (ILC). Even though the $e^{+}e^{-}$
cross sections are below the hadronic cross sections, the strongly
reduced number of background events renders the search easier for
the Higgs-pair signal, through $b\bar b b\bar b$ final states for
instance, in the $e^{+}e^{-}$ environment than in jetty LHC final
states. For sufficiently high luminosities the first phase of
these colliders with an energy of 500 $GeV$ will allow the
experimental analysis of self-couplings for Higgs bosons in the
intermediate mass range. Other couplings between heavy and light
MSSM Higgs bosons can be measured  as well, though only in
restricted areas of the ($\sqrt{s}, \kappa$) plane as illustrated
in the sets of Figs.~\ref{f6} and ~\ref{f9}.

In summary, we have analyzed the triple Higgs bosons self-coupling
at future $e^+e^-$ colliders energies, with the reactions
$e^{+}e^{-}\rightarrow b \bar b hh(HH, AA)$ and
$e^{+}e^{-}\rightarrow t \bar t hh(HH,AA)$. The first is the most
important after considering the $h$ decay, and is statistically
sufficient for an accurate determination of $\kappa$. In the case
of $e^{+}e^{-}\rightarrow t \bar t hh(HH, AA)$, after we consider
the $t$ and $h$ decays the final number of events is small;
although they produce a non-vanishing number of events,
statistically it is insufficient for the determination of
$\kappa$. In these conditions the number of events is small but
our results have never been reported in the literature before and
could be of relevance for the scientific community.

\vspace{1cm}

\begin{center}
{\bf Acknowledgments}
\end{center}

We acknowledge support from SNI and PROMEP (M\'exico). O. A.
Sampayo would like to thank CONICET (Argentina).

\newpage

\vspace{1.5cm}

\begin{references}%{99}


\bibitem{TAbe} American Linear Collider Group (T. Abe {\it et al.}), hep-ex/0106057.

\bibitem{Aguilar} ECFA/DESY Lc Physics Working Group (J. A. Aguilar-Saavedra {\it et al.}), hep-ph/0106315.

\bibitem{Koh} ACFA Linear Collider Working Group (Koh Abe {\it et al.}), hep-ph/0109166.

\bibitem{ILC} ILC Technical Review Committee, second report, 2003, SLAC-R-606, February 2003.

\bibitem{Accomando} CLIC Physics Working Group (E. Accomando {\it et al.}), hep-ph/0412251.

\bibitem{Spira} M. Spira, A. Djouadi, D. Graudenz and P. M. Zerwas, {\it Nucl. Phys. B} {\bf 453}, 17 (1995).

\bibitem{Djouadi1} A. Djouadi, W. Kilian, M. Muhlleitner and P. M. Zerwas, {\it Eur. Phys. J. C.} {\bf 10}, 45 (1999), and references therein.

\bibitem{Weiglein} LHC/LC Study Group (A. G. Akeroyd {\it et al.}), {\it Phys. Rept.} {\bf 426}, 47 (2006), and references therein.

\bibitem{Higgs} P. W. Higgs, {\it Phys. Rev. Lett}. {\bf 13}, 508 (1964).

\bibitem{Higgs1} P. W. Higgs, {\it Phys. Lett.} {\bf 12}, 132 (1964).

\bibitem{Higgs2} P. W. Higgs, {\it Phys. Rev.} {\bf 45}, 1156 (1966).

\bibitem{Englert} F. Englert, and  R. Brout, {\it Phys. Rev. Lett.} {\bf 13}, 321 (1964).

\bibitem{Guralnik} G. S. Guralnik, C. R. Hagen and T. W. B. Kibble, {\it Phys. Rev. Lett.} {\bf 13}, 585 (1964).

\bibitem{Weinberg} S. Weinberg, {\it Phys. Rev. Lett.} {\bf 19}, 1264 (1967).

\bibitem{Salam} A. Salam, in {\it Elementary Particle Theory}, ed. N. Southolm (Almqvist and Wiksell, Stockholm, 1968), p. 367.

\bibitem{Glashow} S.L. Glashow, {\it Nucl. Phys}. {\bf 22}, 579 (1961).

\bibitem{Nilles} H. P. Nilles, {\it Phys. Rep.} {\bf 110}, 1 (1984).

\bibitem{Haber} H. E. Haber and G. L. Kane, {\it Phys. Rep. C} {\bf 117}, 75 (1985).

\bibitem{Barbieri} Riccardo Barbieri, {\it Riv. Nuovo Cimento} {\bf 11}, 1 (1988).

\bibitem{Gunion} J. F. Gunion and H. E. Haber, {\it Nucl. Phys. B} {\bf 272}, 1 (1986).

\bibitem{Gunion1} J. F. Gunion and H. E. Haber, {\it Nucl. Phys. B} {\bf 307}, 445 (1988).

\bibitem{Li} S. P. Li and M. Sher, {\it Phys. Lett. B} {\bf 149}, 339 (1984).

\bibitem{Gunion2} J. F. Gunion and A. Turski, {\it Phys. Rev. D} {\bf 39}, 2701 (1989).

\bibitem{Gunion3} J. F. Gunion and A. Turski, {\it Phys. Rev. D} {\bf 40}, 2325 (1989).

\bibitem{Gunion4} J. F. Gunion and A. Turski, {\it Phys. Rev. D} {\bf 40}, 2333 (1989).

\bibitem{Okada} Yasuhiro Okada, Masahiro Yamaguchi and Tsutomu Yanagida, {\it Prog. Theor. Phys.} {\bf 85}, 1 (1991).

\bibitem{Haber1} H. E. Haber and R. Hempfling, {\it Phys. Rev. Lett.} {\bf 66}, 1815 (1991).

\bibitem{Ellis} John R. Ellis, Giovanni Ridolfi and Fabio Zwirner, {\it Phys. Lett. B} {\bf 257}, 83 (1991).

\bibitem{Dress} M. Dress and M. N. Nojiri, {\it Phys. Rev. D} {\bf 45}, 2482 (1992).

\bibitem{Djouadi2} A. Djouadi, H. E. Haber and P. M. Zerwas, {\it Phys. Lett. B} {\bf 375}, 203 (1996).

\bibitem{Djouadi3} A. Djouadi, W. Kilian, M. M. Muhlleitner and P. M. Zerwas, {\it Eur. Phys. J. C} {\bf 10}, 27 (1999).

\bibitem{Osland0} P. Osland, P. N. Pandita, {\it Phys. Rev. D} {\bf 59}, 055013 (1999).

\bibitem{Boudjema} F. Boudjema and A. Semenov, hep-ph/0201219.

\bibitem{Djouadi4} Abdelhak Djouadi, {\it Phys. Rept.} {\bf 459}, 1 (2008), and references therein.

\bibitem{Ferrera} Giancarlo Ferrera, Jaume Guasch, David L\'opez-Val and Joan Sola, {\it Phys. Lett. B} {\bf 659}, 297 (2008).

\bibitem{Osland} P. Osland and N. Pandita, hep-ph/9911295.

\bibitem{Osland2} P. Osland and N. Pandita, hep-ph/9902270.

\bibitem{Osland1} Per Osland, P. N. Pandita and Levent Selbuz, {\it Phys. Rev. D} {\bf 78}, 015003 (2008).

\bibitem{Abdesslam} Abdesslam Arhrib, Rachid Benbrik and Cheng-Wei Chiang, {\it Phys. Rev. D} {\bf 77}, 115013 (2008).

\bibitem{Lafaye} R. Lafaye, D.J. Miller, M. Muhlleitner and S. Moretti, hep-ph/0002238.

\bibitem{Miller} D. J. Miller and S. Moretti, {\it Eur. Phys. J. C} {\bf 13}, 459 (2000).

\bibitem{Vernon} Vernon Barger, Tao Han, Paul Langacker, Bob McElrath and Peter Zerwas, {\it Phys. Rev. D} {\bf 67}, 115001 (2003).

\bibitem{A.Gutierrez} A. Guti\'errez-Rodr\'{\i}guez, M. A. Hern\'andez-Ru\'{\i}z and O. A. Sampayo,
                      {\it Phys. Rev. D} {\bf 67}, 074018 (2003).

\bibitem{A.Gutierrez1} A. Guti\'errez-Rodr\'{\i}guez, M. A. Hern\'andez-Ru\'{\i}z and O. A. Sampayo, {\it Mod. Phys. Lett. A} {\bf 20}, 2629 (2005).

\bibitem{A.Gutierrez2} C. A. B\'aez, A. Guti\'errez-Rodr\'{\i}guez, M. A. Hern\'andez-Ru\'{\i}z and O. A. Sampayo, {\it Acta Phys. Slov.} {\bf 56}, 455 (2006).

\bibitem{A.Gutierrez3} A. Guti\'errez-Rodr\'{\i}guez, A. del Rio-De Santiago and M. A. Hern\'andez-Ru\'{\i}z,   {\it J. Phys. Conf. Ser.} {\bf 37}, 34 (2006).

\bibitem{A.Gutierrez4} A. Guti\'errez-Rodr\'{\i}guez, M. A. Hern\'andez-Ru\'{\i}z and O. A. Sampayo, A. Chubykalo and A. Espinoza,
                       {\it  AIP Conf. Proc.} {1026}, 269 (2008).

\bibitem{A.Gutierrez5} A. Guti\'errez-Rodr\'{\i}guez, M. A. Hern\'andez-Ru\'{\i}z and O. A. Sampayo, Andrei Chubykalo and A. Espinoza-Garrido,
                      {\it J. Phys. Soc. Jpn.} {\bf 77}, 094101 (2008), arXiv: 0807.0663 [hep-ph].

\bibitem{Pukhov} A.Pukhov, E. Boos, M. Dubinin, V. Edneral, V. Ilyin, D. Kovalenko, A. Kryukov, V. Savrin, S. Shichanin and A. Semenov, hep-ph/9908288.

\bibitem{A.Gutierrez6} A. Guti\'errez-Rodr\'{\i}guez, M. A. Hern\'andez-Ru\'{\i}z and O. A. Sampayo, hep-ph/0601238.


\end{references}
\end{document}